\RequirePackage{ifpdf}
\documentclass[12pt]{article}
\usepackage{epsfig,graphicx,bm,amsmath,amsfonts}
\usepackage{amssymb,xcolor,float}
\usepackage{parskip}

\usepackage[english]{babel}
\usepackage[T1]{fontenc}
\usepackage[utf8]{inputenc}
\usepackage{authblk}

\usepackage{hyperref}
\usepackage[capitalize]{cleveref}
\numberwithin{equation}{section} 

\definecolor{refcol}{rgb}{0.9,0.1,0.1}
\hypersetup{colorlinks=true,linkcolor=blue,citecolor=refcol,urlcolor=black,linktocpage}

\newcommand{\ben}{\begin{eqnarray}\displaystyle}
	\newcommand{\een}{\end{eqnarray}}

\newcommand{\be}{\begin{equation}}
	\newcommand{\ee}{\end{equation}}

\newcommand{\bc}{\begin{center}}
	\newcommand{\ec}{\end{center}}

\newcommand{\eesp}{\end{split}}
\newcommand{\bsp}{\begin{split}}

\newcommand{\cI}{\mathcal{I}}

\newcommand{\cL}{\mathcal{L}}

\newcommand{\cO}{\mathcal{O}}

\newcommand{\cR}{\mathcal{R}}

\newcommand{\cV}{\mathcal{V}}
\newcommand{\cW}{\mathcal{W}}

\newcommand{\ra}{\rightarrow}

\newcommand{\lb}{\left (}
\newcommand{\rb}{\right )}

\newcommand{\bensp}{\begin{eqnarray}\begin{split}}
		\newcommand{\eensp}{\end{eqnarray}\end{split}}

\newcommand{\dow}{\partial}

\newcommand{\vphizero}{  \varphi_{\!{}_\infty} }

\usepackage{wrapfig}

\begin{document}
	
	\parindent=12pt
	\begin{center}
		{\Large \textbf{String Theory Corrections to Holographic Black Hole Chemistry}}
	\end{center}
	
	\baselineskip=18pt
	\bigskip
	\centerline{\textbf{Suvankar Dutta and Gurmeet Singh Punia}}
	\bigskip
	\centerline{\it Department of Physics,}
	\centerline{\it	Indian Institute of Science Education and Research Bhopal,}
	\centerline{\it Bhopal 462066, India}
	\bigskip
	\centerline{E-mail: \href{mailto:suvankar@iiserb.ac.in}{suvankar@iiserb.ac.in}, \; \href{mailto:gurmeet17@iiserb.ac.in}{gurmeet17@iiserb.ac.in}}
	\vskip .6cm
	\medskip
	
	\vspace*{4.0ex}
	
	\centerline{\bf Abstract} \bigskip
	
	\noindent 
	The connection between the bulk and the boundary first law of thermodynamics in adS space has been discussed in generic higher derivative gravity. String theory corrections to supergravity render higher derivative terms in the bulk action, proportional to different powers of string theory parameter $\alpha'$. A variation in the cosmological constant induces a variation in the 't Hooft coupling in the boundary theory. We show that in order to match the bulk first law and Smarr relation with the boundary side we need to include the variation of $\alpha'$ in the bulk thermodynamics as a book keeping device. Accordingly, the boundary first law and Euler relation are modified with the inclusion of two central charges ($a$, $c$) and/or other chemical potentials as thermodynamic variables. We consider four and six derivative terms as well as the $\text{Weyl}^4$ terms (in type IIB) in bulk in support of our generic result.
	
	\vfill\eject

	\tableofcontents
	

	\section{Introduction}\label{sec:intro}
	
	The thermodynamics of anti-de Sitter (adS) black holes entered into a new paradigm after considering the cosmological constant as the thermodynamic pressure and inclusion of its variation in the first law \cite{Kastor:2009wy,Kubiznak:2012wp, Cai:2013qga,  Altamirano:2013uqa, Altamirano:2013ane, Dutta:2013dca, Johnson:2014yja, Kubiznak:2014zwa, Kubiznak:2016qmn}. This new paradigm is dubbed as \emph{black hole chemistry} \cite{Kubiznak:2014zwa, Kubiznak:2016qmn}. In the context of the adS/CFT correspondence the black hole provides a dual description of the field theory living on the boundary, hence it is expected that the thermodynamic variables and the laws on both sides match. In order to make the bulk first law and Smarr relation consistent with the boundary thermodynamics it was shown by \cite{Karch:2015rpa, Visser:2021eqk, Cong:2021jgb} (following the earlier work \cite{Kastor:2010gq}) that the inclusion of variation of Newton's constant along with the cosmological constant in the bulk first law is required. Before we elaborate further let us summarise the current status of the first law and the Smarr relation for uncharged adS-black holes in two derivative gravity\footnote{The formulae can be written for electrically charged black holes as well.}. 
	
	The first law in the bulk is given by
	\begin{eqnarray}\label{eq:bulkfirstlaw}
		dM = \frac{T}{4 G} dA + \frac{\Theta}{8 \pi G} d\Lambda -M \frac{dG}{G}
	\end{eqnarray}
	where
	\begin{eqnarray}
		\begin{aligned}
			M & = \text{ADM mass of the black hole},\quad 
			T  = \text{Hawking temperature},\\
			A & = \text{Area of the event horizon},\quad
			\Lambda  = \text{Cosmological constant},\\
			G & = \text{Newton's constant}.
		\end{aligned}
	\end{eqnarray}
	The parameter $\Theta$ has a geometrical interpretation in terms of proper volume weighted locally by the norm of the Killing vector $\xi$ \cite{Kastor:2009wy, Cvetic:2010jb,Jacobson:2018ahi}
	\begin{eqnarray}\label{eq:thetadef}
		\Theta = \int_{\text{BH}} |\xi| \ dV - \int_{\text{adS}} |\xi|\ dV
	\end{eqnarray}
	where the integrations are taken over the constant time hypersurfaces in black hole and adS spacetimes. The generalised Smarr relation for $d+1$ dimensional bulk is given by \cite{Smarr:1972kt, Banerjee:2010ye}
	\begin{eqnarray}\label{eq:smarr}
		M = \frac{d-1}{d-2} \frac{T A}{4 G} - \frac{1}{d-2}\frac{\Theta \Lambda}{4 \pi G}.
	\end{eqnarray}
	
	The inclusion of variation of Newton's constant in the first law is needed in order to make the black hole thermodynamics consistent with boundary one. Without including the variation of $G$ and considering the bulk pressure $P= - \Lambda/8\pi G$, the first law (\ref{eq:bulkfirstlaw}) takes the form 
	\ben\label{eq:bulkfirstlawnoG}
	dM = T dS + \cV dP 
	\een 
	with $\cV=-\Theta$ and the Smarr relation becomes $M = (d-1)/(d-2) TS - 2/(d-2) P\cV$. There are two main objections with (\ref{eq:bulkfirstlawnoG}), in the context of the adS/CFT correspondence. From a simple computation of asymptotic stress energy tensor \cite{Balasubramanian:1999re} one can find both the energy and the pressure of the boundary theory. The boundary pressure evaluated in this way does not match with the bulk pressure defined above. Also the energy of the boundary theory turns out to be equal to the ADM mass $M$ of the black hole. Whereas $M$, that appears in the first law (\ref{eq:bulkfirstlawnoG}) is identified with the enthalpy of the black hole system not energy. Hence, we see that the bulk first law (\ref{eq:bulkfirstlawnoG}) can not be interpreted as the boundary first law.
	
	In adS$_{d+1}/CFT_d$ correspondence the adS radius $L$, the effective $d+1$ dimensional Newton's constant $G$ and the number of colours $N$ (number of coincident $D$ branes) are related by
	\begin{equation}\label{eq:dictionary1}
		\text{dictionary 1 :}\quad  \frac{L^{d-1}}{G} \sim N^2.
	\end{equation}
	The exact relation depends on the particular string theory and its reduction over the compact manifold. Since the bulk cosmological constant is given by
	\begin{equation}\label{eq:Lambdadef}
		\Lambda = - \frac{d(d-1)}{2L^2},
	\end{equation}
	a variation of $\Lambda$ in the bulk, therefore, induces a variation in the number of colour $N$ (degrees of freedom or the central charge) in the boundary. Also, from the asymptotic structure of the adS metric it follows that the spatial volume of the boundary theory goes like $V \sim L^{d-1}$. As a result $\Lambda$ variation also induces a variation in the spatial volume of the boundary theory. In order to disentangle the variations of $N$ and $V$ on the boundary (as a result of $\Lambda$ variation) the variation of the bulk Newton's constant was included in the bulk thermodynamics as a ``book keeping" device \cite{Karch:2015rpa}, such that a variation in $N$ at fixed $V$ in the boundary corresponds to variation of $G^{-1}$ at fixed $L$ in the bulk and variation of $V$ at fixed $N$ corresponds to variation of $L$ keeping $L^{d-1}/G$ fixed in the bulk. After the inclusion of $G$ variation, the bulk first law takes the form (\ref{eq:bulkfirstlaw}).
	
	Inclusion of $G$ variation in the first law has an advantage that the first law (\ref{eq:bulkfirstlaw}) can be re-written in the following way
	\begin{equation}\label{eq:bulkfirstlaw2}
		dM = T dS - \frac{M}{d-1}\frac{d L^{d-1}}{L^{d-1}} + (M- T S)\frac{d(L^{d-1}/G)}{L^{d-1}/G}
	\end{equation}
	and hence can immediately be mapped to the boundary first law \cite{Visser:2021eqk, Cong:2021jgb, Cong:2021fnf}. $L^{d-1}$ is proportional to the thermodynamic volume of the boundary theory. The coefficient of 
	$d L^{d-1}$	term is therefore identified with the pressure of the boundary theory which satisfies the equation of state : $E = M = (d-1) p V$. Finally, the last term in the first law $d(L^{d-1}/G)/(L^{d-1}/G)$ is identified with the variation of the central charge $c$ and its coefficient is a new chemical potential $\mu_c$. This new chemical potential  satisfies the boundary Euler relation
	\begin{equation}\label{eq:euler}
		E =M = TS + \mu_c c.
	\end{equation}
	
	In this paper we establish the connection between the bulk and the boundary thermodynamics in adS space in generic higher derivative gravity\footnote{Large $N$ corrections to holographic Smarr relation in presence of Lovelock gravity was considered in \cite{Sinamuli:2017rhp}. }. String theory correction to supergravity renders higher derivative terms in the bulk action, proportional to different powers of string theory parameter $\alpha'$. A priori it may seem trivial to extend the connection beyond the supergravity limit by including higher derivative corrections to all the thermodynamic variables. But one has to be careful because the second holographic dictionary (\ref{eq:dictionary2}) implies that the variation of the cosmological constant in the bulk also induces a variation of the 't Hooft coupling $\lambda$ of the boundary theory. In order to disentangle the variation of $\lambda$ from the variation of $N$ and volume $V$, one needs to include the variation of $\alpha'$ (along with $G$ and $L$) in the bulk as a book keeping device. We show that under a suitable change of thermodynamic variables the bulk first law can be interpreted as the boundary first law and the Smarr relation renders the generic Euler relation of the boundary theory. 
	
	The summary of our results is as follows. We find that the variation of $G$ and $\alpha'$ in the bulk first law can be traded with the variations of two boundary central charges and thus the boundary first law can be written as
	\begin{equation}\label{eq:bulkfirstlawb}
		dM_c = T_c dS_c -\frac{M_c}{d-1} \frac{dL_c^{d-1}}{L_c^{d-1}} + \mu_+ dc_+ +\mu_- dc_-.
	\end{equation}
	The subscript $c$ denotes the higher derivative corrected thermodynamic variables. $c_\pm$ are related to two boundary central charges $c$ and $a$ as $c_\pm = (c\pm a)/2$, $\mu_\pm$ are the corresponding chemical potentials (associated with $c_\pm$ respectively) satisfying the generic Euler relation
	\begin{equation}\label{eq:eulerc}
		E_c = M_c = T_c S_c + \mu_+ c_+ + \mu_- c_-.
	\end{equation}
	
	The organisation of this paper is following. In section \ref{sec:bulklaws} we discuss how bulk first law and Smarr relation are modified in presence of generic higher derivative terms. In the next section (sec \ref{sec:holofirstllaw}) we show that how the bulk first law can be interpreted as the boundary first law in generic higher derivative gravity. In section \ref{sec:examples} we discuss few examples in support of our generic statement. In particular we consider six derivative corrections as well as $Weyl^4$ correction to supergravity action.
	
	\section{The bulk first law and Smarr relation in higher derivative gravity}\label{sec:bulklaws}
	
	In the throat limit \cite{Maldacena:1997re} the effective $d+1$ dimensional action has the following qualitative form\footnote{In writing (\ref{eq:actionhd}) we have made another simplification. In principle there could be other matter terms also, but we have ignored those terms.} (in the Einstein frame)
	
	\begin{equation}\label{eq:actionhd}
		\cI = \frac{1}{16 \pi G} \int d^{d+1}x \sqrt{-g} \lb R - 2\Lambda + \sum_{n\geq 2} (\alpha')^{n-1}\cR^{(2n)}\rb
	\end{equation}
	where $\alpha'$ is proportional to the square of string length. $\cR^{(2n)}$ is the $2n$ derivative term in the action. 
	
	Such higher derivative terms appear in low energy effective action of different closed
	string theories. For example the appearance of curvature square term ($Riemann^2 \sim \cR^{(4)}$) in
	heterotic string theory is well known. $\cR^{(8)}$ terms appear in superstring theories whereas $\cR^{(6)}$ appears in bosonic string theory. In general such lower dimensional effective actions are also endowed with other fields, for example $U(1)$ gauge fields, higher form fields, dilaton etc. However, in this section we shall consider the effect of pure curvature higher derivative terms on the boundary thermodynamics. In section \ref{sec:W4} we include the effect of dilaton coupled to higher derivative terms in the bulk and boundary thermodynamics.
	
	The supergravity limit corresponds to $\alpha'/L^2 \ll 1$. In this limit all the higher derivative terms drop out. The structures of these higher derivative terms are completely fixed for a specific string theory. The only parameter which appears in front of these terms is different powers of $\alpha'$ as this is the only dimensional full parameter in the theory. In the context of gauge/gravity duality such higher derivative terms correspond to the large 't Hooft coupling correction in the strongly correlated field theory. The adS/CFT correspondence provides another relation between the parameters in the string theory and gauge theory
	\ben\label{eq:dictionary2}
	\text{dictionary 2 : } \alpha' = \frac{L^2}{\sqrt{\lambda}}
	\een
	where $\lambda = N g_s$ is the 't Hooft coupling of the boundary theory and $g_s$ is the string coupling. This means that the two derivative gravity (supergravity) is dual to strongly coupled gauge theory on the boundary. 
	
	From the relation (\ref{eq:dictionary2}) we see that a variation of $\Lambda$ induces a variation in the 't Hooft coupling constant $\lambda$. Therefore, as before, to disentangle the $\lambda$ variation from $N$ and $V$ variations on the boundary side we allow the parameter $\alpha'$ to vary in the bulk along with $L$ and $G$ as a book keeping device \cite{Karch:2015rpa,Cong:2021fnf}. A variation with respect to $\lambda$ on the boundary corresponds to variation of $\alpha'$ keeping other combinations fixed.
	
	The Smarr relation for Lovelock gravity in adS spacetime was considered in \cite{Kastor:2010gq, Sinamuli:2017rhp}. Considering the variations of the coefficients of Lovelock terms they derived the Smarr relation and showed that it gives the mass of the black hole in terms of geometrical quantities together with the parameters of the Lovelock theory. Following the similar argument the Smarr relation in a generic higher derivative gravity (\ref{eq:actionhd}) takes the following form
	\ben\label{eq:smarrc}
	M_c = \frac{d-1}{d-2} \frac{T_c A_c}{4 G} - \frac{1}{d-2}\frac{\Theta_c \Lambda_c}{4 \pi G} + \frac{2}{d-2}  \frac{U_{\alpha'}}{G} \alpha'
	\een
	and the first law turns out to be
	\begin{eqnarray}\label{eq:bulkfirstlawc}
		dM_c = \frac{T_c}{4 G} dA_c + \frac{\Theta_{c}}{8 \pi G} d\Lambda_c - M_c \frac{dG}{G} + \frac{U_{\alpha'}}{G} d\alpha'.
	\end{eqnarray}
	In presence of higher derivative terms the mass, entropy, temperature and cosmological constant receive corrections. Here we denote the higher derivative corrected thermodynamic quantities with the same variables with a subscript $c$. The new thermodynamic variable $U_{\alpha'}$ is conjugate to the coupling constant $\alpha'$ and $U_{\alpha'}d\alpha'$ term in the first law disappears in the supergravity limit : $\alpha'=0$. The quantity $A_c$ appearing in the first law is given by
	\begin{equation}\label{eq:Ac}
		A_c = 4 G S_c.
	\end{equation}
	We call this quantity the ``Wald area". Later we shall see that in our parametrisation the horizon area remains unchanged under higher derivative corrections. The Wald area is equal to the horizon area in the $\alpha'\ra 0$ limit. The Wald area will appear in the first law of black hole thermodynamics. In presence of higher derivative terms the effective radius of the adS spacetime changes. Denoting the effective radius by $L_c$, the corrected cosmological constant $\Lambda_c$ is given by $\Lambda_c = -d(d-1)/2L_c^2$. The value of $L_c$ depends on the nature of higher derivative terms.
	
	The first law and Smarr relation are consistent with those given in \cite{Kastor:2010gq}. However, unlike \cite{Kastor:2010gq} we have only one coupling constant $\alpha'$. Another important point to note here is that unlike two derivative gravity the variable $\Theta_c$ does not have the geometrical meaning (\ref{eq:thetadef}) any more. In section \ref{sec:sixderi} and \ref{sec:W4} we explicitly calculate all the higher derivative corrected thermodynamic variables up to order $(\alpha')^2$ and $(\alpha')^3$ respectively and show that the first law and Smarr relation are satisfied.
	
	Our next goal is to use the adS/CFT dictionary (\ref{eq:dictionary1}, \ref{eq:dictionary2}) to obtain the boundary first law from the bulk first law by suitably choosing boundary thermodynamic variables. We also show that the bulk Smarr relation boils down to the generic Euler relation under that choice.

	
	\section{Holographic first law and Euler equation} \label{sec:holofirstllaw}
	
	The first law written in terms of $(A_c,\Lambda_c,G,\alpha')$ can not immediately be identified with the boundary first law. In order to do so we need to express the first law in terms of boundary thermodynamic variables. In two derivative gravity it was shown that instead of $(A, \Lambda, G)$ one can write the first law in $(S,L^{d-1},c)$ basis, where $L^{d-1}$ is (proportional to) spacial volume and $c$ is the central charge of the boundary theory \cite{Visser:2021eqk, Cong:2021fnf, Cong:2021jgb}. In higher derivative gravity we have an extra variable $\alpha'$ in the bulk. Therefore the natural question is what is the correct thermodynamic basis in the boundary theory in this case. 
	
	CFTs in higher dimensions are endowed with two central charges $c$ and $a$. These two central charges are an artifact of breaking of conformal symmetry at quantum level. The expectation value of trace of CFT stress tensor is given by $\langle T^\mu_\nu\rangle = -a E_4 -c I_4$ where $E_4$ and $I_4$ are two invariants made of Riemann tensor. The holographic computation of these central charges \cite{Henningson:1998gx} shows that $c=a \sim L^{d-1}/G$ in the supergravity limit. However, in the presence of string theory corrections they are not the same anymore, they differ by the inverse powers of the 't Hooft coupling \cite{Banerjee:2009fm, Nojiri:1999mh, Blau:1999vz}. Motivated by \cite{Visser:2021eqk, Cong:2021jgb, Cong:2021fnf} we identify that the extra bulk parameter $\alpha'$ can be replaced in terms of the second central charge $a$. However, instead of writing the first law in terms of $(c, a)$ we define a new set
	\begin{equation}\label{eq:cpm}
		c_\pm = \frac{c \pm a}{2}
	\end{equation}
	and write down the first law in the $(c_+,c_-)$ basis such that in the $\alpha'\ra 0$ limit we readily get back the two derivative results.
	
	The holographic dictionary (\ref{eq:dictionary2}) relates two dimension full parameters $\alpha'$ and $L$ in the bulk with a dimensionless parameter $\lambda$ on the boundary. The effective (corrected) length of adS spacetime therefore can be written as,
	\begin{equation}
		L_c = L \tilde b(\lambda)
	\end{equation}
	where $\tilde b(\lambda)$ depends on the form of the higher derivative terms and $\tilde b(\lambda) =1$ as $\lambda\ra \infty$. Therefore the string theory parameter $\alpha'$ can be written in terms of the effective radius of adS spacetime as
	\begin{equation}\label{eq:alphaLc}
		\alpha' = \frac{L_c^2}{\sqrt{\lambda}\ \tilde b^2(\lambda)} = L_c^2 b(\lambda), \quad \text{where} \quad b(\lambda) = \frac{1}{\sqrt{\lambda}\ \tilde  b^2(\lambda)}.
	\end{equation}
	From the similar dimensional analysis the generic form of $c_\pm$ in higher derivative gravity can be written as,
	\begin{equation}\label{eq:cpmhpm}
		c_+= \frac{L_c^{d-1}}{G} h_+(\lambda), \quad \text{and} \quad c_- = \frac{L_c^{d-1}}{G} h_-(\lambda)
	\end{equation}
	where $h_+$ and $h_-$ are functions of dimensionless parameter $\lambda$ and depend on the nature of the higher derivative terms added. In holographic theory they also satisfy
	\begin{eqnarray}
		h_+(\lambda) \sim 1
		\quad \text{and} \quad h_{-}(\lambda) \sim \frac{1}{\sqrt{\lambda}} \quad \lambda \ra \infty.
	\end{eqnarray}
	
	Varying equations (\ref{eq:cpmhpm}) we find
	\begin{equation}\label{eq:dcpm}
		dc_\pm = c_\pm \frac{dL_c^3}{L_c^3} - c_\pm \frac{dG}{G} + \frac{c_\pm h'_\pm}{h_\pm} d\lambda. 
	\end{equation}
	Taking variation of equation (\ref{eq:alphaLc})
	\begin{equation}
		d\alpha' = \frac{2 b L_c^2}{3} \frac{dL_c^3}{L_c^3}  + L_c^2 b' d\lambda
	\end{equation}
	we replace $d\alpha'$ in terms of $\frac{dL_c^3}{L_c^3}$ and $d\lambda$ in the bulk first law. We then solve (\ref{eq:dcpm}) to replace $dG$ and $d\lambda$ in the first law in terms of $dc_\pm$. We also use the Smarr relation (\ref{eq:smarrc}) to replace $\Theta_c$ in the first law. After simplification the final result is given by
	\begin{equation}
		\begin{aligned}
			dM_c = T_c dS_c - \frac{M_c}{(d-1) L_c^{d-1}} dL_c^{d-1} & + \lb\frac{h_+' \left(M_c-S_c T_c\right)-c_- U b' L_c^{3-d}}{ \left(c_+ h_-'-c_-
				h_+'\right)}\rb dc_+ \\
			& + \lb \frac{c_+ U b' L_c^{3-d} -  h_+' \left(M_c-S_c T_c\right)}{\left(c_+ h_-'-c_-
				h_+'\right)}\rb dc_-.
		\end{aligned}
	\end{equation}
	Identifying the coefficient of $dc_\pm$ as the chemical potentials $\mu_\pm$ associated with $c_\pm$ we see that $\mu_\pm$ satisfy the generic Euler relation (\ref{eq:eulerc}).
	
	Thus we see that the bulk first law (\ref{eq:bulkfirstlawc}) can immediately be identified with extended first law of the boundary CFT (\ref{eq:bulkfirstlawb}) and the generic Smarr relation renders the Euler relation (\ref{eq:eulerc}). As a consistency check we see that the chemical potentials $c_- = 0$, $c_+ = c =a$ in the limit $\lambda \ra \infty$ and we get back (\ref{eq:bulkfirstlaw2}) and (\ref{eq:euler}). We also note that though individual thermodynamic quantities explicitly depend on specific combinations of higher derivative terms, the first law and the Euler relation are independent of any specific combinations.
	
	In the next section we consider a few examples of higher derivative gravity and compute different thermodynamic quantities and chemical potential in order to check our generic statement.


	\section{Higher derivative thermodynamics : examples}
	\label{sec:examples}
	
	The higher derivative terms which appear in the effective $d+1$ dimensional Lagrangian under a consistent truncation of string theory have a very specific form. For example if we consider type $IIB$ string theory on adS$_5\times S^5$ background and truncate the string theory action over $S^5$ the resulting effective theory in adS$_5$ has the first non-trivial higher derivative correction term at the order of $(\alpha')^3$ \cite{Grisaru:1986vi,Freeman:1986zh,Gross:1986iv,Banks:1998nr,Pawelczyk:1998pb}. We shall consider the effect of such terms on bulk and boundary thermodynamics in sec. \ref{sec:W4}. Before that, we consider a bulk action with four and six derivative terms. The curvature square terms appear in heterotic string theory whereas the curvature cube terms appear in bosonic string theory. The four derivative terms in the action are proportional to $\alpha'$ and the six derivative terms appear at the order $(\alpha')^2$. Our goal is to compute all the thermodynamic quantities up to order $(\alpha')^2$ and check the generic first law, Smarr relation and the Euler relation. In this section we work in $4+1$ dimensions.
	
	\subsection{Six derivative theory}
	\label{sec:sixderi}
	
	We start with the most general four and six derivative terms. Before we find the corrections to bulk metric and thermodynamic quantities we briefly discuss the field re-definition ambiguity with these terms. There are five possible dimension-6 invariants which do not involve Ricci tensors or curvature scalars,
	\ben
	\begin{aligned}
		I_1 &=
		R^{\mu\nu}_{\, \, \, \, \alpha\beta}R^{\alpha\beta}_{\, \, \, \,
			\lambda\rho}R^{\lambda\rho}_{\, \, \, \, \mu\nu},\quad 
		I_2 =
		R^{\mu\nu}_{\, \,\, \, \rho\sigma}R^{\rho\tau}_{\, \, \, \, \lambda\mu} R^{\sigma \ \ \lambda}_{\ \ \tau \ \ \nu}, \quad 
		I_3 =
		R^{\alpha\nu}_{\, \,\, \, \mu\beta}R^{\beta\gamma}_{\, \, \, \, \nu\lambda}
		R^{\lambda\mu}_{\, \,\,\, \gamma\alpha},
		\\
		I_4 &= R_{\mu\nu\alpha\beta}R^{\mu \alpha}_{\, \, \, \, \gamma\delta}
		R^{\nu\beta\gamma\delta},\quad 
		I_5 = R_{\mu\nu\alpha\beta}{\cal D}^2 R^{\mu\nu\alpha\beta}\ .
	\end{aligned}
	\een
	These five invariants satisfy the following relations,
	\be
	I_3= I_2-{\frac{1}{4}}I_1, \quad I_4={\frac{1}{2}}I_1, \quad I_5 = -I_1 - 4 I_2 \ .
	\ee
	Hence only two of them are independent.  We choose these two invariants to be $I_1$ and $I_2$.
	
	Including the invariants made out of Ricci tensor and scalar the most general Lagrangian (density) containing all possible independent curvature invariants are given by
	\ben \label{6dacn}
	\begin{aligned}
		\cL  = &  a_0 R - 2 \Lambda  +
		\alpha'  \bigg (  \beta_1 R^2 + \beta_2
		R_{\mu\nu\rho\sigma} R^{\mu\nu\rho\sigma} + \beta_3 R_{\mu\nu}R^{\mu\nu}  \bigg )  \\
		& + \alpha'^2 \bigg ( \alpha_1 I_1 + \alpha_2 I_2 + \alpha_3
		R_{\mu\alpha\beta\gamma}R_{\nu}^{\,\,\alpha\beta\gamma}R^{\mu\nu} + \alpha_4
		R R_{\mu\nu\rho\sigma}R^{\mu\nu\rho\sigma} + \alpha_5
		R_{\mu\nu\rho\lambda} R^{\nu\lambda}R^{\mu \rho}
		\\
		& + \alpha_6 R_{\mu \nu} R^{\nu
			\lambda} R^{\mu}_{\, \, \lambda}+ \alpha_7 R_{\mu\nu}{\cal D}^2 R^{\mu\nu}  + \alpha_8 R R_{\mu\nu}R^{\mu\nu} + \alpha_9
		R^3 + \alpha_{10} R {\cal D}^2 R \bigg )+ {\cal O}(\alpha'^3)\ . 
	\end{aligned}
	\een
	However, many terms in this Lagrangian are ambiguous up to a field re-definition \cite{Metsaev:1986yb, Banerjee:2009fm}. Under the following field re-definition
	\ben \label{redef1}
	\begin{aligned}
		g_{\mu \nu} \ra {\tilde g}_{\mu\nu} = g_{\mu \nu}
		& + \alpha' \big (d_1 g_{\mu\nu} R + d_2 R_{\mu\nu}
		\big ) + \alpha'^2 \big (d_3 R_{\mu \alpha \beta \gamma}R_{\nu}^{\, \, \alpha \beta
			\gamma} + d_4 g_{\mu\nu}
		R_{\alpha\beta\gamma\sigma} R^{\alpha\beta\gamma\sigma}
		\\
		& 
		+
		d_5 R_{\mu \alpha\beta\nu}R^{\alpha\beta}  + d_6 R_{\mu \lambda}R^{\lambda}_{\,
			\, \nu} + d_7 {\cal D}^2 R_{\mu\nu}
		+ d_8 g_{\mu\nu}
		R_{\alpha\beta}R^{\alpha\beta}  \\
		&  + d_9 g_{\mu\nu} R^2 + d_{10} g_{\mu\nu} {\cal
			D}^2 R \big ) + {\cal O}(\alpha'^3)
	\end{aligned}
	\een
	only the coefficients $a_0,\beta_2, \alpha_1$ and $\alpha_2$ remain invariant as it is not possible to generate any higher rank tensor from a lower rank tensor in (\ref{redef1}). Therefore the coefficients $\beta_2, \alpha_1$ and $\alpha_2$ are unambiguous. By proper choice of field re-definition one can set all other ambiguous coefficients to zero
	\ben \label{6dacnspe}
	\ {\cal L} &\ra&  a_0 R - 2 \Lambda +
	\alpha'  \ \beta_2
	R_{\mu\nu\rho\sigma} R^{\mu\nu\rho\sigma}
	\ + \alpha'^2 \bigg ( \alpha_1 I_1 + \alpha_2 I_2 \bigg).
	\een
	In the Euclidean approach the thermodynamic quantities are obtained by evaluating the Euclidean on-shell action. Since the action is not invariant under such field re-definition the thermodynamic quantities depend on un-ambiguous as well as different ambiguous coefficients. However, in this section we turn on only the un-ambiguous higher derivative terms i.e. $\beta_2$, $\alpha_1$ and $\alpha_2$ to maintain the simplicity of different thermodynamic variables. In the appendix we present the results for other ambiguous coefficients also.

	\subsubsection{The solution}\label{sec:solution}
	
	We start with the following 6-derivative action
	\ben \label{eq:6dacnreqg}
	\begin{aligned}
		{\cal I}  = \frac{1}{16\pi G} \int & d^5x \sqrt{-g} \bigg [ R - 2 \Lambda +
		\alpha'\   \beta_2
		R_{\mu\nu\rho\sigma} R^{\mu\nu\rho\sigma} + \alpha'^2 \bigg ( \alpha_1 I_1 + \alpha_2 I_2 \bigg) \bigg]\ .
	\end{aligned}
	\een
	The coefficients $\beta_2$, $\alpha_1$ and $\alpha_2$ are fixed, they do not vary.
	
	The equations of motion for the metric obtained from the action \eqref{eq:6dacnreqg} are given by
	\begin{align}\label{eq:eom}
		R_{\alpha \beta} -\frac{1}{2} R g_{\alpha \beta} - \frac{6}{L^{2}} g_{\alpha \beta} = \alpha^\prime T^{(4)}_{\alpha \beta} + {\alpha^\prime}^2 T^{(6)}_{\alpha \beta}
	\end{align}
	where $ T^{(4)}_{\alpha \beta} $  and $ T^{(6)}_{\alpha \beta} $ are given by \eqref{eq:4deom} and \eqref{eq:6deom}, respectively. In order to solve this equations we consider the following metric ansatz (for a static spherically symmetric solution)
	\begin{equation}\label{eq:metricansatz}
		ds^2 = - f(r) dt^2 + \frac{1}{g(r)} dr^2 + r^2 d\Omega^2_3 \, ,
	\end{equation}
	where $  d\Omega^2_3 $ is a metric on a 3-sphere sphere of unit radius. We solve the Einstein's equations perturbatively to obtain $f(r)$ and $g(r)$. In absence of any higher derivative terms the equations of motion admit an asymptotically adS black hole solution given by\footnote{The subscript $0$ means the leading solution.}
	\begin{equation}\label{eq:metric0}
		f_0(r) = {g_0(r)} = -\frac{\left(r^2-r_+^2\right) \left(L^2+r^2+r_+^2\right)}{L^2 r^2}
	\end{equation}
	where $ r_+ $ is the horizon radius which is related to the ADM mass by \cite{Hawking:1982dh} 
	\begin{equation}\label{eq:admmass0}
		M = \frac{3 \pi  r_+^2 \left(L^2+r_+^2\right)}{8 G L^2}.
	\end{equation} 
	
	Treating the higher derivative terms perturbatively one can systematically find the corrections to the leading solutions. The equations of motion (\ref{eq:eom}) are ordinary second order differential equations (while treated perturbatively). Therefore at every order in $\alpha'$ we have two integration constants. We fix these constants by demanding that the horizon radius $r_+$ remains unaffected under perturbations and the spacetime remains adS in the limit $r\ra \infty$. With these two conditions the final form of $f(r)$ and $g(r)$ are given appendix \ref{app:metric}. Asymptotically the metric takes the following form
	\begin{equation}\label{eq:adsmetbound}
		ds^2 \sim -\left(1 + \frac{r^2}{L_{c}^2} \right)dt^2 + \left(1 + \frac{r^2}{L_{c}^2} \right)^{-1} dr^2 + r^2 d\Omega^2_3 \, ,
	\end{equation} 
	where
	\ben\label{eq:Lc}
	\begin{aligned}
		L_c  = L \Bigg[ 1-\frac{\beta_2}{3} \frac{\alpha '}{L^2} & -\frac{1}{6} \left(4
		\alpha _1+3 \alpha _2 + \frac{5}{3}\beta_2^2 \right) \left(\frac{\alpha '}{L^2}\right)^2 \Bigg]
	\end{aligned}
	\een
	is the corrected adS radius. For a consistency, one can also compute the Ricci scalar for the corrected solution (\ref{eq:fc}, \ref{eq:gc}) and check that
	\begin{equation}\label{eq:ricci}
		R = - \frac{12}{L_c^2}.
	\end{equation}
	
	\subsubsection{The thermodynamic variables }\label{sec:thermodynamics}
	
	Once we have the black hole metric corrected up to $(\alpha')^2$ one can compute different thermodynamic variables associated with the corrected geometry.
	
	The correction to the black hole temperature can be computed in different ways. One simple method to compute is the Euclidean method. In this method we first Euclideanise the time direction by replacing $t\ra i \tau$. The Euclidean metric will show a conical singularity at $r=r_+$ unless the Euclidean time $\tau$ is periodic. One can compute the periodicity $\beta$ of $\tau$ and the black hole temperature is inversely proportional to the periodicity
	\begin{equation}\label{hawkingtempdef}
		T = \frac{1}{\beta} = \frac{1}{4\pi} \sqrt{g'(r)f'(r)} \bigg|_{r_+}.
	\end{equation}
	After simplification we find that the corrected temperature is given by
	\begin{align}\label{eq:temp}
		\begin{aligned}
			T_c = &  \left(\frac{r_+}{\pi L^2}+\frac{1}{2\pi r_+}\right)- \lb \frac{2  \beta _2
				\left(3 L^4+6 L^2 r_+^2+2 r_+^4\right)}{3 \pi  L^2 r_+^3} \rb \frac{\alpha'}{L^2} \\
			& + \bigg ( \frac{2 \left(-21 L^6+18 L^4 r_+^2+99 L^2 r_+^4+62 r_+^6\right)\alpha_1}{3 \pi  L^2 r_+^5} + \frac{(L^6-3 r_+^2 \left(2 L^2+r_+^2\right) \left(L^2+2 r_+^2\right))\alpha_2}{2 \pi  L^2 r_+^5} \\
			& \quad +  \frac{8 \left(36 L^6+36 L^4 r_+^2-39 L^2 r_+^4-38 r_+^6\right)\beta_2^2}{9 \pi  L^2 r_+^5}\bigg)\lb \frac{\alpha'}{L^2}\rb^2.
		\end{aligned}
	\end{align}
	
	The entropy of the black hole can also be computed either using Wald's formula or Euclidean method. In either method the corrected entropy turns out to be
	\begin{eqnarray}\label{eq:Waldent}
		\begin{aligned}
			S_c =  &\frac{\pi ^2 r_+^3}{2 G} \bigg[1 +   \beta _2\bigg( \frac{12 L^2}{r_+^2} + 8  \bigg) \frac{\alpha^\prime}{L^2} + \bigg( \frac{12 L^2 \left(12 \alpha _1+3 \alpha _2 - 24 \beta _2^2\right)}{r_+^2} \\
			& + \frac{3 L^4 \left(36 \alpha _1+3 \alpha _2 - 48 \beta _2^2 \right)}{r_+^4} + \frac{36 \left(4 \alpha _1+3 \alpha _2\right) - 416 \beta_2^2}{3}\bigg) \left( \frac{\alpha'}{L^2}\right)^2\bigg].
		\end{aligned}
	\end{eqnarray}
	
	The ADM mass of the black hole can also be obtained by either computing the asymptotic stress tensor \cite{PhysRev.116.1322,Balasubramanian:1999re} or on-shell Euclidean action \cite{Hawking:1982dh,Dutta:2006vs}. The result is given by
	\begin{eqnarray}\label{eq:adm}
		\begin{aligned}
			M_c  = &  \frac{3 \pi}{8 \pi G} \bigg[ r_+^2 + \frac{r_+^4}{L^2} + \beta_2 \bigg( \frac{6 L^4 + 20r_+^4 + 24 r_+^2 L^2}{3 L^2} \bigg) \frac{\alpha^\prime}{L^2} + \bigg( \frac{2L^4 \left(32 \alpha _1+\alpha _2 - 48 \beta _2^2 \right)}{r_+^2}\\
			& + 3 \lb 84 \alpha _1 + 5 \alpha _2 - 128 \beta _2^2\rb {L^2} + \frac{r_+^2}{3}(3 \left(276 \alpha _1+45 \alpha _2 \right)
			-1424 \beta _2^2) \\
			& + \frac{r_+^4 }{9 L^2} (3 \left(268 \alpha _1+99 \alpha _2\right) -1648 \beta _2^2) \bigg) \left( \frac{\alpha'}{L^2}\right)^2 \bigg].
		\end{aligned}
	\end{eqnarray}
	
	\subsubsection{The first law and chemical potentials}\label{sec:6dfirstlaw}
	
	In order to write down the higher derivative corrected first law of thermodynamics we first note that the effective radius of adS spacetime has been modified (\ref{eq:Lc}). As a result the cosmological constant $\Lambda$ will also be corrected 
	\begin{equation}
		\Lambda \ra \Lambda_c = - \frac{6}{L_c^2}.
	\end{equation}
	Allowing the variations of $G$, $L$ and $\alpha'$ we find that the thermodynamic variables $M_c$, $T_c$, $A_c$ and $\Lambda_c$ satisfy the first law (\ref{eq:bulkfirstlawc}). The thermodynamic potentials $\Theta_c$ and $U_{\alpha'}$ are given by
	\begin{eqnarray}\label{eq:Thetac}
		\begin{aligned}
			\Theta_c & =     -\frac{1}{2} \pi ^2 r_+^4-\frac{2 \pi ^2 \beta _2 r_+^2 \alpha ' \left(2
				L^2+r_+^2\right)}{3 L^2} + \frac{\pi^2}{18} \bigg(-36 \alpha _1
			\left(39 L^4+74 L^2 r_+^2+35 r_+^4\right)\\ 
			& + 27 \alpha _2 \left(3 L^4+2 L^2 r_+^2 - r_+^4 \right) + 16 \beta _2^2 \left(96 L^4 + 190 L^2 r_+^2 + 93 r_+^4\right) \bigg) \lb \frac{\alpha'}{L^2}\rb^2
		\end{aligned}
	\end{eqnarray}
	and 
	\begin{align}\label{eq:Ualpha}
		\begin{aligned}
			U_{\alpha'} & = -\frac{\pi \left(9 L^4+20 L^2 r_+^2+6 r_+^4\right) \beta _2 }{4L^4} + \frac{\pi}{12 r_+^2 L^4} \bigg( 12 \alpha _1 \left( - 6 L^6 + 9 L^4 r_+^2 + 39 L^2 r_+^4 + 19 r_+^6 \right)\\
			& -9 \alpha _2 \left(4 L^6+21 L^4 r_+^2+27 L^2 r_+^4 + 15 r_+^6\right) + 16 \beta _2^2 \left(9 L^6 + 24 L^4 r_+^2+19 L^2 r_+^4+5 r_+^6\right) \bigg)\frac{\alpha'}{L^2}.
		\end{aligned}
	\end{align}
	As we mentioned in the introduction, the thermodynamic potential $\Theta_c$ associated with $\Lambda_c$ variation does not have the geometrical meaning (\ref{eq:thetadef}) in presence of higher derivative terms. For the spacetime metric (\ref{eq:fc}, \ref{eq:gc}) the correction to the geometric volume $\Theta_c$ is different from (\ref{eq:Thetac}). With this corrected thermodynamic potentials it is easy to check that the corrected Smarr relation (\ref{eq:smarrc}) is satisfied.
	
	To cast the bulk first law in terms of boundary variables we compute the anomaly coefficients $c$ and $a$ in presence of higher derivative terms in the action (\ref{eq:6dacnreqg}). The answers are given by \cite{Banerjee:2009fm}
	\begin{eqnarray}
		\begin{aligned}
			c  = \frac{L_c^3}{128 \pi G}\bigg[1 & + \frac{4 \beta_2}{\sqrt{\lambda}} +\frac{ \left( - 36 \alpha _1 + 21 \alpha
				_2\right)}{\lambda}\bigg]
		\end{aligned}
	\end{eqnarray}
	and
	\begin{eqnarray}
		\begin{aligned}
			a = \frac{L_c^3}{128 \pi G}\bigg[ 1 & - \frac{4 \beta _2}{\sqrt{\lambda}} + \frac{12 \alpha _1 + 9 \alpha_2}{\lambda }\bigg].
		\end{aligned}
	\end{eqnarray}
	Replacing the variations $dG$ and $d\alpha'$ as discussed in sec. \ref{sec:holofirstllaw}, in terms of $dc_\pm$, we can write the first law in the form given in (\ref{eq:bulkfirstlawb}) with
	\begin{equation}\label{eq:mup6d}
		\begin{aligned}
			\mu_+ & = \frac{16 \pi ^2 r_+^2 \left(r_+^2-L^2\right)}{L^5} - \frac{128 \pi ^2 \beta _2 
				\left(3 L^4 + 6 L^2 r_+^2 + 2 r_+^4\right)}{3 \sqrt{\lambda } L^5} \\
			& -\frac{8 \pi ^2}{9 \lambda  L^5 r_+^2} \bigg(36 L^6 \left(40 \alpha _1+5 \alpha _2-64 \beta _2^2\right) + 36 L^4
			r_+^2 \left( 12 \alpha _1 + 24 \alpha _2 - 61 \beta _2^2\right)\\
			& + 3 L^2 r_+^4 \left( - 1200 \alpha
			_1 + 396 \alpha _2 + 875 \beta _2^2\right)+r_+^6 \left(-2688 \alpha _1+432 \alpha _2+2423
			\beta _2^2\right)\bigg)
		\end{aligned}
	\end{equation}
	and\footnote{Note that $\beta_2$ is a non-zero coefficient. It comes in the denominator because $c_- \sim \beta_2$.}
	\begin{align}\label{eq:mum6d}
		\begin{aligned}
			\mu_- & = \frac{8 \pi ^2 \left(9 L^4+20 L^2 r_+^2+7 r_+^4\right)}{L^5} +\frac{8 \pi ^2}{3 \beta _2 \sqrt{\lambda } L^5 r_+^2} \Big(24 \left(L^2+r_+^2\right) \Big( 3 L^4 (2 \alpha _1+\alpha _2) \\
			& + 6 L^2 r_+^2 (2 \alpha _1+\alpha _2) + r_+^4 (4 \alpha _1 + 3 \alpha _2)\Big) -\beta _2^2 \left(288 L^6 + 756 L^4 r_+^2 + 591 L^2 r_+^4 + 147 r_+^6\right)\Big)
		\end{aligned}
	\end{align}
	and they satisfy Euler relation (\ref{eq:eulerc}) up to order $\frac1\lambda$.
	
	\subsection{\texorpdfstring{$\mathcal{W}^4$}{} term}\label{sec:W4}
	
	Finally, we discuss a string theory example in the context of the adS/CFT.  Since the conjecture is valid for the complete string theory, one should consider the stringy corrections to the 10D supergravity action. In particular we consider string theory correction to type IIB supergravity. The first corrections occur at order $(\alpha')^3$ \cite{Grisaru:1986vi,Freeman:1986zh, Gross:1986iv}. The bosonic part of the action in the Einstein frame is given by
	\begin{equation}
		\label{eq:W4action}
		I = \frac{1}{16 \pi G_{10}} \int d^{10} x \; \sqrt{-g} \Big\{ R - \frac{1}{2} \left( \partial \phi \right)^2 - \frac{1}{4.5!} \frac{1}{N^2} F^2_5 + \gamma e^{-\frac{3}{2} \phi} \cW^4 \Big\}
	\end{equation}
	where $ F_5 $ a self dual 5-form field strength, $\phi$ is the dilaton and $ \mathcal{W}^4 $ denotes the \emph{eight derivative} term in action, which can be expressed as contraction of four Weyl tensors
	\begin{equation}\label{eq:W4}
		\mathcal{W}^4 = C_{a b c d} C^{e b c f} C^{a}{ }_{g h e} C_{f}^{g h d}+\frac{1}{2} C_{a d b c} C^{e f b c} C^{a}{ }_{g h e} C_{f}^{g h d}.
	\end{equation}
	The coupling constant $\gamma$ is given by
	\begin{equation}\label{eq:gammaalpha}
		\gamma = \frac{1}{8} \zeta(3) (\alpha')^{3} .
	\end{equation}
	These higher derivative terms do not alter the extremal adS$_5\times S^5$ geometry \cite{Banks:1998nr,Kallosh:1998qs}. However this observation is not true in the non-extremal case \cite{Pawelczyk:1998pb,Landsteiner:1999gb}. Moreover the higher derivative action (\ref{eq:W4action}) is not supersymmetrically complete. Supersymmetric completion of eight derivative terms in type IIB string theory can be found in \cite{deHaro:2002vk, Green:2003an,Peeters:2003ys,Peeters:2003pv, Rajaraman:2005up, Policastro:2006vt,Paulos:2008tn}. A tentative corrections to black hole free energy and other thermodynamic quantities in presence of these supersymmetrically complete terms were considered in \cite{Paulos:2008tn}. The answer is tentative in a sense that the author did not compute the corrections to the full black hole geometry in presence of these terms since those computations are extremely cumbersome. In this paper we shall consider only the $\cW^4$ term (\ref{eq:W4}) in the action and see how the bulk and boundary first laws are modified.
	
	In the supergravity limit ($\gamma \ra 0$) the type IIB vacuum admits a solution of the form $\text{adS}_5 \times S^5$ with constant five form field strength over $\text{adS}_5$ and $S^5$ with constant dilaton. As a result one can truncate the 10 dimensional action over $S^5$ and the effective five dimensional action takes the form of Einstein-Hilbert action in $\text{adS}_5$. For $\cW^4$ term apparently it appears that one can proceed exactly in the similar way what we discussed in the previous sections i.e. replacing the $dG$ and $d\alpha'$ terms in the bulk first law in terms of the variations of two central charges $c_\pm$. However, it turns out that the central charges $a$ and $c$ do not receive any corrections in presence of $\cW^4$ terms and hence $c=a +\cO(\gamma^2)$. Therefore the question is what is the relevant boundary parameter with which we trade the $d\alpha'$ term in the bulk first law.
	
	The dilaton field plays an important role here. The massless dilaton field in $4+1$ dimensions corresponds to a dimension 4 scalar operator $\hat \cO_4 \sim \frac{1}{g_{YM}^2} \int \text{Tr}F^2$ in the boundary theory. The dilaton has the asymptotic fall-off
	\ben
	\phi(r) = \vphizero + \frac{\varphi_1}{r^4} + \cO(r^{-6})
	\een
	where $\vphizero$ plays the role of source for $\hat \cO_4$ and $\varphi_1$ is the expectation value of $\hat \cO_4$ \cite{Balasubramanian:1998sn,Klebanov:1999tb}. At the leading order ($\gamma =0$) the dilaton is constant and the expectation value of $\hat \cO_4$ is zero. However at sub-leading order $\vphizero$ induces an expectation value for $\cO_4$ proportional to $\gamma$. Hence the boundary first law and Euler relation can be written in terms of $\langle\cO_4\rangle$ and other standard thermodynamic variables.
	
	\subsubsection{\texorpdfstring{$\mathcal{W}^4$}{} corrected geometry and the boundary first law}
	
	Following \cite{Pawelczyk:1998pb,Landsteiner:1999gb} we consider the following ansatz for the ten dimensional metric and the five form field strength
	\begin{equation}
		d s^{2}=\frac{r^{2}}{L^{2}} e^{-\frac{10}{3} C(r)}\left(e^{2 A(r)+8 B(r)} d \tau^{2}+e^{2 B(r)} d r^{2} + L^2 d \Omega_{3}^{2}\right)+e^{2 C(r)} L^2 d \Omega_{5}^{2}
	\end{equation}
	\begin{align}
		F_5 = \mathcal{F}_5 + \star \mathcal{F}_5 \,, \quad \mathcal{F}_5 = dB_4
	\end{align}
	where 
	\begin{equation}
		B_4 = f(r) \, dt \wedge d\emph{vol}_3.
	\end{equation}
	Here $d\emph{vol}_3$ be the volume element of $S^3$ with radius $L$. The leading order metric solution is given by (subscript $0$ stands for leading solution)
	\begin{equation}
		\begin{aligned}
			&A_0(r)=-2 \log \left(\frac{r}{L}\right)+\frac{5}{2} \log \left(\frac{r^{2}}{L^{2}}+\frac{r^{4}}{L^{4}}-\frac{r_{0}^{2}}{L^{4}}\right) ; \\
			&B_0(r)=-\frac{1}{2} \log \left(\frac{r^{2}}{L^{2}}+\frac{r^{4}}{L^{4}}-\frac{r_{0}^{2}}{L^{4}}\right) ; \\
			&C_0(r)=0 .
		\end{aligned}
	\end{equation}
	The leading dilaton field is constant and denoted as $\vphizero$ and the solution of four form field is $B_4 = \sqrt{2} \frac{r^4}{L^4} \, dt \wedge d\emph{vol}_3 $. Here $ r_0 $ is the non-extremality parameter i.e. in the limit $r_0\ra 0$ the ten dimensional geometry corresponds to the near horizon limit of an extremal $D3$ branes. We consider the higher derivative terms perturbatively, such that $ A(r) = A_0(r) + \gamma A_1(r) $, $B(r) = B_0(r) + \gamma B_1(r)$, $f(r) = f_0(r)+\gamma f_1(r)$ and $\varphi = \vphizero + \gamma \phi_1$. Solving the ten dimensional equations of motion up to $\cO(\gamma)$ we find that the corrections are given by 
	\begin{align}
		\begin{aligned}
			A_1(r) = e^{-\tfrac{3 \vphizero }{2}} (r^4 + L^2 \left(r^2-r_0^2\right))^{-1} \left( \frac{5 r_0^6 \left( 312 r^4 + L^2 (272 r^2 - 237 r_0^2) \right)}{2 r^{12}} \right. \\
			\left. -\frac{5 r_0^6 \left(312 r_+^4 + L^2 \left(272 r_+^2-237 r_0^2\right) \right)}{2 r_+^{12}} \right),
		\end{aligned}\\
		\begin{aligned}
			B_1(r) = e^{-\tfrac{3 \vphizero }{2}} (r^4 + L^2 \left(r^2-r_0^2\right))^{-1} \left( \frac{r_0^6 \left(312 r_+^4 + L^2 \left(272 r_+^2-237 r_0^2\right) \right)}{2 r_+^{12}} \right. \\
			\left. -\frac{5 r_0^6 \left(64 L^2 r^2-57 L^2 r_0^2+72 r^4\right)}{2 r^{12}} \right),
		\end{aligned}
	\end{align}
	\begin{equation}
		\begin{aligned}
			f_1(r) = 60 \sqrt{2} e^{-\tfrac{3}{2} \vphizero} \frac{r_0^6}{ L^4 r^8}.
		\end{aligned}
	\end{equation}
	Here $r_+$ is the corrected horizon radius. Correction to the the dilaton solution is given by
	\begin{equation}
		\begin{aligned}
			\phi_1(r) = 45 e^{-\frac{3}{2} \vphizero } \frac{\left(L^8+4 L^6 r_+^2+7 L^4 r_+^4+6 L^2 r_+^6+3 r_+^8\right)}{4 L^4 r_+^4 \left(L^2+r_+^2\right){}^3}  \log \left(\frac{L^2+r^2+r_+^2}{r^2}\right) \\
			-\frac{e^{-\frac{3}{2} \vphizero}}{16 L^4 r^{12} r_0^4} \left(36 L^4 r_0^8 r^2+30 L^4 r_0^{10}+60 L^2 r_0^4 r^6 \left(L^2+2 r_0^2\right) + 45 L^4 r_0^6 r^4 \right. \\
			\left. + 45 L^4 r_0^8 r^4 +180 r^{10} \left(L^4+4 L^2 r_0^2+3 r_0^4\right) + 90 r_0^2 r^8 \left(L^4+3 L^2 r_0^2+r_0^4\right)\right).
		\end{aligned}
	\end{equation}
	The asymptotic expansion of dilaton field near AdS boundary is given by
	\begin{equation}
		\phi(r) = \vphizero - \frac{45 \gamma  e^{-\frac{3}{2} \vphizero} \left(L^2+r_+^2\right){}^4}{8 L^6 r_+^4} \frac{1}{r^4} + \mathcal{O}\left(\frac{1}{r^6}\right).
	\end{equation}
	Following \cite{Balasubramanian:1998sn,Klebanov:1999tb} we find the expectation value of the corresponding boundary operator $\cO_4$ is given by
	\begin{equation}
		\langle\hat \cO_4 \rangle = 4 \varphi_1 = - \frac{45 \gamma  e^{-\frac{3}{2} \vphizero} \left(L^2+r_+^2\right){}^4 }{2 L^6 r_+^4}.
	\end{equation}
	
	From the asymptotic expansion of the metric one can easily show that unlike four and six derivative case, the adS radius $L$ does not receive any higher derivative correction.
	
	\subsubsection{Bulk thermodynamics and the first law}	
	Using the Euclidean technique \eqref{hawkingtempdef} one can correct the higher derivative correction to the black hole temperature and it is given by is (also  using the the relation (\ref{eq:gammaalpha}))
	\begin{equation}
		T_c = \frac{L^2+2 r_+^2}{2 \pi  L^2 r_+} \left( 1  -\frac{5 \zeta (3) e^{-\frac32 \varphi _{\infty }}  \left(L^2-3 r_+^2\right) \left(L^2+r_+^2\right){}^3 \left(\alpha '\right)^3
		}{4 L^6 r_+^6 \left(L^2+2 r_+^2\right)}\right).
	\end{equation}
	Correction to the entropy can be computed using the Wald formula or Euclidean method. The corrected entropy is given by,  
	\begin{equation}
		S_c = \frac{\pi^2 {r_+}^3}{2 G} \left( 1 + \frac{15 \zeta (3) e^{-\frac{3}2 \varphi_\infty} \left(L^2+r_+^2\right){}^3 \left(\alpha '\right)^3}{2 L^6
			r_+^6} \right).
	\end{equation}
	ADM mass of the black hole is given by 
	\begin{equation}
		M_c = \frac{3 \pi  r_+^2 \left(L^2+r_+^2\right)}{8 G L^2} \left( 1 + \frac{5 \zeta (3) e^{-\frac32 \varphi_\infty}  
			\left(L^2+r_+^2\right){}^2 \left(7
			L^2+15 r_+^2\right) \left(\alpha '\right)^3 }{8 L^6 r_+^6} \right). 
	\end{equation}
	
	In presence of moduli the first law of black hole changes \cite{Gibbons:1996af}. Including the variation of $\vphizero$ in the first law we find that
	\begin{equation}\label{eq:W4bulkfirstlaw}
		dM_c = \frac{T_c}{4 G} dA_c + \frac{\Theta_{c}}{8 \pi G} d \Lambda_c - \frac{M_c}{G} dG + U_{\alpha^\prime} d (\alpha^\prime)^3 + \mu_\varphi d\vphizero 
	\end{equation}
	where 
	\begin{align}
		\Theta_c & = -\frac{ \pi ^2 r_+^4}{2} - \frac{15 \pi ^2 \zeta (3) e^{-\frac32 \varphi_\infty} \left(L^2+r_+^2\right){}^3 \left(\alpha
			'\right)^3
		}{4
			L^6 r_+^2}\,, \\
		U_{\alpha^\prime} & = - \frac{15 \pi \zeta(3) e^{-\frac{3}{2} \vphizero } \left(L^2+r_+^2\right){}^4}{64 G L^8 r_+^4} \, , \\
		\mu_\varphi & = \lb\frac{\dow M_c}{\dow\vphizero}\rb = - \frac{\pi}{8 G L^2} \langle  \hat{\mathcal{O}_4}\rangle \,.
	\end{align}
	and these thermodynamics quantities satisfying the Smarr relation
	\begin{equation}\label{eq:W4smarr}
		M_c = \frac{3}{2} T_c S_c + \frac{\Theta_{c}}{8 \pi G} \Lambda_c + 3 \frac{U_{\alpha^\prime}}{G}\lb \alpha' \rb^3 .
	\end{equation} 
	
	\subsubsection{Boundary thermodynamics}
	
	The dilaton couples to the bulk operator $\frac{1}{g_{YM}^2}\text{Tr}F^2$. We define a variable
	\be\label{eq:psidef}
	\psi = e^{-\varphi_\infty}
	\ee
	and use the holographic dictionary
	\begin{equation}\label{eq:dictionary3}
		\alpha^\prime \sim \frac{L^2}{\sqrt{\lambda}}, \quad \lambda \sim \frac{L^{3/2} e^{\varphi_\infty}}{\sqrt{G}}  \quad \text{and} \quad c = c_+ \sim \frac{L^3}{G}
	\end{equation}
	to replace variations of $\alpha'$, $\varphi_\infty$ and $G$ in terms of variation of $c_+$ and $\psi$ in the bulk first law (\ref{eq:W4bulkfirstlaw}) to obtain the same in terms of boundary parameters
	\begin{equation}\label{eq:W4boundaryfl}
		dE_c=dM_c = T_c dS_c - \frac{M_c}{3} \frac{d{L_c^3}}{L_c^3} + \mu_c d c - \mu_{\psi} d \psi.
	\end{equation}
	where $L_c=L$ and 
	\begin{eqnarray}
		\begin{aligned}
			\mu_c & = \frac{16 \pi ^2 r_+^2 \left(L^2-r_+^2\right)}{L^5} + \frac{5 \pi ^2 \zeta (3) e^{-3 \varphi /2} \left(13 L^2-51 r_+^2\right) \left(L^2+r_+^2\right){}^3 \lb\alpha'\rb^3}{2 L^{11} r_+^4}\\
			\mu_\psi & = -  \frac{2 \mu_\varphi}{\psi}  .
		\end{aligned}
	\end{eqnarray}
	Here we see that the pressure satisfies the equation of state $E_c = 3 p V$. The chemical potentials $ \mu_c $ and $\mu_\psi$ satisfy the Euler equation 
	\begin{equation}\label{eq:eulerW4}
		E_c = T_cS_c + \mu_c c + \frac{1}{4} \mu_\psi \psi .
	\end{equation}  
	
	\section{Summary and Discussion}\label{sec:dis}
	
	In this paper we consider string theory corrections to black hole thermodynamics in adS space and its consistency with the thermodynamics of the boundary theory in the context of the adS/CFT correspondence. Since the higher derivative terms are low energy effects of some bona fide string theories their couplings are also fixed. The only parameter that appears in the bulk action with these higher derivative terms is $\alpha'$. From the adS/CFT dictionary we see that a variation of $\Lambda$ induces a variation in the ’t Hooft coupling $\lambda$, apart from variations in colour $N$ and boundary volume $V$. Therefore, to disentangle the $\lambda$ variation from that of $N$ and $V$ we allow the parameter $\alpha'$ to vary in the bulk along with $L$ and $G$ as a book keeping device. This allows us to establish the equivalence between the bulk and boundary thermodynamics. We consider two types of examples. In the first type we added pure metric higher derivative terms in the action (for example four and six derivative terms). In presence of such terms we include the variation of $\alpha'$ in the bulk first law and show that trading the variations of $G$ and $\alpha'$ with the variations of $c_+$ and $c_-$, where $c_\pm = (c\pm a)/2$, the bulk first law can be beautifully interpreted as the boundary first law which is written in terms of variations of $c_\pm$. As a result the boundary theory is endowed with two chemical potentials $\mu_\pm$ (corresponding to $c_\pm$) and they satisfy the generalised Euler relation (\ref{eq:eulerc}). In the second example we considered an eight derivative term in the bulk Lagrangian coming from the superstring theory. In this case the term is coupled with the dilaton. In the leading case the dilaton solution is constant and we see that the effective five dimensional bulk first law is the same as before. In presence of the higher derivative term the dilaton solution is modified and it turns out that the dilaton sources an expectation value of a dimension $4$ operator namely $\text{Tr}F^2$ and the expectation value is $\sim \alpha'^3$. In this case we trade the $G$ and $\alpha'$ variations in the bulk with the variations of $c$ and the asymptotic value of the dilaton, which acts as a source for the $\text{Tr}F^2$ operator and write the boundary first law in terms of their variations. Again the boundary theory is endowed with two chemical potential $\mu_c$ corresponds to $c$ and $\mu_\varphi$, proportional to $\langle \text{Tr}F^2 \rangle$, corresponds to $\varphi_\infty$. These two chemical potentials satisfy the generalised Euler relation (\ref{eq:eulerW4}).
	
	The phase structure of adS black holes in higher derivative gravity is endowed with an extra chemical potential and hence the dimension of the thermodynamic phase phase will increase. In this paper we study the thermodynamics perturbatively, however it would be interesting to study the black hole phase structure in presence of the extra parameter, even perturbatively. It would also be interesting to find an effective Van-der-Waals type description (following\cite{Dutta:2021whz}) of higher derivative black holes and understand the effect of the central charges on the mean-field potential.\\
	
	\centerline{***************}
	
	\vspace{1cm}
	\noindent
	{\bf Acknowledgement :} We are thankful to Arnab Rudra, Muktajyoti Saha and Archit Vidyarthi for their helpful discussion. The work of SD is supported by the MATRICS (grant no. MTR/2019/ 000390, the Department of Science and Technology, Government of India). Finally, we are indebted to people of India for their unconditional support toward the researches in basic science.
	\vspace{1cm}
	
	\appendix

	\section{EOM's for 4-derivative and 6-derivative terms}
	
	Here we present the expressions for $T^{(4)}_{ab}$ and $T^{(6)}_{ab}$ appear in the equations of motion \eqref{eq:eom} in presence of four and six derivative terms.
	
	The contribution from the 4-derivative term \textit{i.e.} $R^{\alpha\beta\gamma\delta} R_{\alpha\beta\gamma\delta}$ is given by 
	\begin{equation}\label{eq:4deom}
		\begin{aligned}
			T^{(4)}_{\alpha \beta} =  \frac{\beta_2}{16 \pi G} \Big( 4 R^{\gamma \delta } R_{\alpha \gamma \beta \delta } + 2 R_{\alpha }{}^{\gamma \delta \zeta } R_{\beta \gamma \delta \zeta } - 4 R_{\alpha }{}^{\gamma } R_{\beta \gamma } - \tfrac{1}{2} g_{\alpha \beta } R_{\gamma \delta \zeta \lambda } R^{\gamma \delta \zeta \lambda } \\
			- 2 \nabla_{\beta }\nabla_{\alpha }R + 4 \Box R_{\alpha \beta } \Big)
		\end{aligned}
	\end{equation}
	and the same for the 6-derivative terms is given by
	\begin{equation}\label{eq:6deom}
		T^{(6)}_{\alpha\beta} = \frac{1}{16 \pi G} \left( {T^{(a)}}_{\alpha \beta} + {T^{(b)}}_{\alpha \beta} \right)
	\end{equation}
	where
	\begin{equation}
		\begin{aligned}
			{T^{(a)}}_{\alpha \beta} =&  \alpha_1 \Big( 6 R^{\gamma \delta } \bigl(2 R_{\alpha }{}^{\zeta }{}_{\gamma }{}^{\lambda } R_{\beta \delta \zeta \lambda } + R_{\alpha \gamma }{}^{\zeta \lambda } (R_{\beta \delta \zeta \lambda } + 2 R_{\beta \zeta \delta \lambda })\bigr) - 6 R_{(\alpha|}{}^{\gamma } R_{|\beta)}{}^{\delta \zeta \lambda } R_{\gamma \zeta \delta \lambda }  \\
			& - 3 R_{\alpha }{}^{\gamma \delta \zeta } (R_{\beta }{}^{\mu }{}_{\gamma }{}^{\nu} R_{\delta \zeta \mu \nu} + 4 R_{\beta }{}^{\mu }{}_{\delta }{}^{\nu} R_{\gamma \mu \zeta \nu}) + \tfrac{3}{2} R_{\alpha }{}^{\gamma \delta \zeta } R_{\beta [\gamma|}{}^{\mu \nu} R_{|\delta] \zeta \mu \nu} -  \tfrac{1}{2} g_{\alpha \beta } I_1 \\
			&  + 6 \nabla^{\delta }R_{\alpha }{}^{\gamma } \nabla_{[\delta|}R_{\beta |\gamma] } + 6 R_{(\alpha|\delta \gamma \zeta } \nabla^{\zeta }\nabla^{\delta }R_{|\beta) }{}^{\gamma } + 6 \nabla_{\gamma }R_{\beta \lambda \delta \zeta } \nabla^{\lambda }R_{\alpha }{}^{\gamma \delta \zeta } \Big)
		\end{aligned}
	\end{equation}
	and
	\begin{equation}
		\begin{aligned}
			T^{(b)}_{\alpha \beta} = & \frac{\alpha_2}{2} \Big(6 R^{\gamma \delta } (- R_{\gamma }{}^{\zeta } R_{\alpha \delta \beta \zeta } - R_{\alpha }{}^{\zeta }{}_{\gamma }{}^{\lambda } R_{\beta \lambda \delta \zeta } + R_{\alpha }{}^{\zeta }{}_{\beta }{}^{\lambda } R_{\gamma \zeta \delta \lambda }) + 9 R_{\alpha }{}^{\gamma \delta \zeta } R_{\beta \delta }{}^{\mu \nu} R_{\gamma \mu \zeta \nu}  \\
			&+ 3 R_{\alpha }{}^{\gamma \delta \zeta } R_{\beta }{}^{\mu }{}_{\delta }{}^{\nu} R_{\gamma (\mu| \zeta |\nu)} + 3 R_{\alpha }{}^{\gamma \delta \zeta } R_{\beta }{}^{\lambda }{}_{\gamma }{}^{\lambda 1} R_{\delta \lambda \zeta \lambda 1} -  g_{\alpha \beta } I_2 - 3 \nabla_{\alpha }R^{\gamma \delta } \nabla_{[\beta|}R_{\gamma |\delta]} \\
			& + 3 \nabla^{\delta }R_{\alpha }{}^{\gamma } \nabla_{(\beta |}R_{\gamma |\delta )} - 3 R_{\alpha \gamma \beta \delta } \nabla^{\delta }\nabla^{\gamma }R  + 3 R_{(\alpha| \gamma \delta \zeta } \nabla^{\zeta }\nabla_{|\beta) }R^{\gamma \delta }  \\
			& + 6 R_{\alpha \gamma \beta \delta } \Box R^{\gamma \delta } - 3 \nabla_{\delta }R_{\alpha (\gamma| \beta |\zeta)} \nabla^{\zeta }R^{\gamma \delta } + 12 \nabla_{\zeta }R_{\alpha \gamma \beta \delta } \nabla^{\zeta }R^{\gamma \delta } \\
			&- 3 R_{(\alpha| \gamma \delta \zeta } \nabla^{\zeta }\nabla^{\delta }R_{|\beta )}{}^{\gamma } + 6 R^{\gamma \delta \zeta \lambda } \nabla_{\lambda }\nabla_{\delta }R_{\alpha \gamma \beta \zeta } - 6 \nabla_{\zeta }R_{\beta \delta \gamma \lambda } \nabla^{\lambda }R_{\alpha }{}^{\gamma \delta \zeta } \Big)\,.
		\end{aligned}
	\end{equation}
	
	\section{ \texorpdfstring{$\alpha'$}{txt} corrected metric} \label{app:metric}
	
	In this appendix we present the higher derivative corrected metric in presence of generic four and six derivative terms. Although in our calculations in the main text only three un-ambiguous terms are turned on, but in general one can study black hole thermodynamics in presence of these generic terms. Therefore, it will be helpful to find the perturbative metric up to $\cO(\alpha'^2)$ in presence of all these terms.
	
	Einstein-Hilbert action with negative cosmological constant and generic four-derivative \& six-derivative terms is given by
	\begin{equation}\label{action6derg}
		\begin{aligned}
			I = \frac{1}{16 \pi G } & \int d^5 x \; \sqrt{-g} \; \bigg\{ R - 2 \Lambda  + \alpha'  \bigg (  \beta_1 R^2 + \beta_2 R_{\mu\nu\rho\sigma} R^{\mu\nu\rho\sigma} + \beta_3 R_{\mu\nu}R^{\mu\nu}  \bigg )  \\
			& + \alpha'^2 \bigg ( \alpha_1 {R^{\mu\nu}}_{\alpha\beta} {R^{\alpha\beta}}_{\lambda\rho} {R^{\lambda\rho}}_{\mu\nu} + \alpha_2 {R^{\mu\nu}}_{\rho\sigma} {R^{\rho\tau}}_{\lambda\mu} R^{\sigma \ \ \lambda}_{\ \ \tau \ \ \nu} + \alpha_3 R_{\mu\alpha\beta\gamma}R_{\nu}^{\,\,\alpha\beta\gamma}R^{\mu\nu} \\
			& \qquad \qquad  + \alpha_4 R R_{\mu\nu\rho\sigma}R^{\mu\nu\rho\sigma} + \alpha_5 R_{\mu\nu\rho\lambda} R^{\nu\lambda}R^{\mu \rho} + \alpha_6 R_{\mu \nu} R^{\nu \lambda} R^{\mu}_{\, \, \lambda} \\
			& \qquad \qquad + \alpha_7 R_{\mu\nu}{\cal D}^2 R^{\mu\nu} + \alpha_8 R R_{\mu\nu}R^{\mu\nu} + \alpha_9 R^3 + \alpha_{10} R {\cal D}^2 R \bigg ) \bigg\} \,.
		\end{aligned}
	\end{equation}
	With the metric ansatz \eqref{eq:metricansatz}, the higher derivative corrected metric solution for action \eqref{action6derg} is given by 
	\begin{equation}\label{eq:fc}
		\begin{aligned}
			f(r) =&  1 + \frac{r^2}{L^2}-\frac{r_0^2}{r^2} + \frac{(\alpha ^\prime/L^2)}{3 L^2 r^6} \bigg( 6 \beta _2 L^4 r_0^4+2 \left(10 \beta _1+\beta _2+2 \beta _3\right) r^8 \bigg) \\
			& - \frac{(\alpha ^\prime/L^2)^2}{9 L^2 r^{10}} \bigg[ 72 L^6 r_0^6 \left(\alpha _1+2 \alpha _2-2 \alpha _3-12 \alpha _4+\beta _2 \left(8 \beta _1-7 \beta _2\right)\right)\\
			& \quad - 18 L^6 r_0^4 r^2 \left(36 \alpha _1+9 \alpha _2-24 \alpha _4-16 \beta _2 \left(-\beta _1+5 \beta _2+\beta _3\right)\right) \\
			& \quad - 9 L^4 r_0^4 r^4 \left(132 \alpha _1+27 \alpha _2+8 \alpha _3-40 \alpha _4 + 8 \beta _2 \left(10 \beta _1-33 \beta _2-6 \beta _3\right)\right)\\
			& \quad - r^{12} \Big(3 (4 \alpha _1+3 \alpha _2+8 \alpha _3+40 \alpha _4+16 \alpha _5+16 \alpha _6+80 \alpha _8+400 \alpha _9) \\
			& \qquad +8 \left(10 \beta _1+\beta _2+2 \beta _3\right){}^2\Big) \bigg]
		\end{aligned}
	\end{equation}
	and 
	\begin{equation}\label{eq:gc}
		\begin{aligned}
			g(r) = 1 & +\frac{r^2}{L^2}-\frac{r_0^2}{r^2} + \frac{(\alpha ^\prime/L^2)}{3 L^2 r^{6}} \bigg( 6 \beta _2 L^6 r_0^4+2 \left(10 \beta _1+\beta _2+2 \beta _3\right) L^2 r^8 \bigg) \\
			& + \frac{(\alpha ^\prime/L^2)^2}{9 L^2 r^{10}} \bigg[ 144 L^6 r_0^4 r^2 \left(24 \alpha _1+7 \alpha _3+24 \alpha _4-4 \beta _2 \left(4 \beta _1+8 \beta _2+3 \beta _3\right)\right) \\
			& +18 L^6 r_0^6 \left(-160 \alpha _1+\alpha _2-48 \alpha _3-168 \alpha _4+4 \beta _2 \left(28 \beta _1+51 \beta _2+20 \beta _3\right)\right)\\
			& +9 L^4 r_0^4 r^4 \left(444 \alpha _1+9 \alpha _2+120 \alpha _3+392 \alpha _4-8 \beta _2 \left(26 \beta _1+77 \beta _2+26 \beta _3\right)\right) \\
			& +r^{12} \Big(12 \alpha _1+9 \alpha _2 + 8 \big(3 \alpha _3+15 \alpha _4+6 \alpha _5+6 \alpha _6+30 \alpha _8+150 \alpha _9 \\
			& \qquad +\left(10 \beta _1+\beta _2+2 \beta _3\right){}^2\big) \Big) \bigg]\,.
		\end{aligned}
	\end{equation}
	Here $ r_0 $ is the integration constant and related to the mass of the black hole and $r_+$ be the event horizon $f(r_+)=0$. $ r_0 = 0 $ corresponds to pure adS spacetime solution.

	\section{Euclidean Formulation}
	
	In this appendix we give a quick review of the Euclidean approach to calculating the total energy, entropy and other thermodynamic variables of the adS black holes. We start with the Lorentzian metric that describes pure adS spacetime; after the Wick rotation, $\tau = i \, t$, the metric becomes Euclidean \textit{i.e.} positive definite, where we can construct a thermal state in adS space where the imaginary time coordinate is periodic. The period of the $\tau$ direction is mapped to the inverse temperature of thermal adS gas. The Euclidean metric of $ \text{AdS}_5 $ is given by 
	\begin{equation}
		ds^2 = \left(1+\frac{r^2}{L^2}\right) d\tau^2 + \frac{dr^2}{\left(1+\frac{r^2}{L^2}\right)} + r^2 d\Omega_3^2 \, ,
	\end{equation}
	where $ L $ is the radius of AdS space. Similarly, the Euclidean adS$_5$ Schwarzschild metric is given by
	\begin{equation}
		ds^2 = \left( 1+ \frac{r^2}{L^2} - \frac{r_0^2}{r^2}\right) d\tau^2 + \left( 1+ \frac{r^2}{L^2} - \frac{r_0^2}{r^2}\right)^{-1} dr^2 + r^2 d\Omega_3^2 \,.
	\end{equation}
	where $ r_0 $ be the black hole parameter related with ADM mass of the black hole given as $ M = \tfrac{3 \Omega_3}{16 \pi G} r_0^2 $. This space time has a horizon at $ r = r_+ $, given by the relation $ 1+ \frac{r_+^2}{L^2} - \frac{r_0^2}{r_+^2} = 0$. The Euclidean black hole metric has a conical singularity at $r=r_+$ unless we consider the imaginary time direction to be periodic. This fixes the temperature of the adS Schwarzschild spacetime.
	
	In order to discuss the thermodynamics of the black hole in Euclidean framework \cite{Hawking:1982dh} we define
	the canonical partition function as the functional integration of Euclidean action $I_E$, $ Z = \int [\mathcal{D}g] e^{-I_E} $. In the semi-classical limit $G\ra 0$, the predominant contributions to the path integral comes from the classical solution. So $ \ln \mathcal{Z} = - I_E^{OS} $, where $ I_E^{OS} $ be the on-shall action. Thus free energy of the system is
	\begin{equation}\label{freeE}
		\log Z = - I_E = -\beta F \, ,
	\end{equation}
	and the thermodynamical energy (or ADM mass) of the black hole is
	\begin{equation}\label{energy}
		E = -\frac{\partial(\log Z(\beta))}{\partial \beta} = \frac{\partial I_E}{\partial \beta},
	\end{equation}
	and the entropy is given by
	\begin{equation}\label{entropy}
		S = \beta \frac{\partial I_E}{\partial \beta} - I_E .
	\end{equation}
	However the above thermodynamic quantities receive divergence contributions since the black hole spacetime has infinite volume (the integration over the radial direction ranges to $\infty$). In order to read off the finite values of the thermodynamic quantities we subtract the contribution of thermal adS spacetime from the black hole as a regularisation prescription.  First, we evaluate the action integral for the black hole putting a cut-off on the radial integration : $r_+ \leq r \leq R_c$ where $R_c$ is an IR cutoff on the spacetime and $ r_+ $ is the outermost horizon. In the pure AdS spacetime the region of integration is $ 0 \leq r \leq R_c$. The important point here is that the temperature of the pure adS spacetime can not be fixed from the conical singularity of the metric as the adS metric is well defined everywhere between $0\leq r< \infty$. Rather we demand that both the adS and black hole spacetimes have the same geometry asymptotically. This fixes the temperature of the adS spacetime
	\begin{equation}\label{eq:betarel}
		\beta_{BH} \sqrt{g_{\tau \tau}^{B H}(r=R_c)} = \beta_{AdS}\sqrt{g_{\tau \tau}^{A d S}(r=R_c)} \,.
	\end{equation}
	Via this consideration, the periodicity of the reference adS spacetime depends on the black hole parameters parameters (such as the mass, temperature $ ( 1/\beta_{BH} )$ etc.).

	\subsection{6-derivative gravity}
	
	Here we present the higher derivative corrected (for the action \eqref{action6derg}) thermodynamic quantities using the Euclidean method.
	
	The temperature of the black hole can be computed from the Euclidean metric using (\ref{hawkingtempdef}) where $ \beta_{BH} $ be the inverse Hawking temperature in corrected geometry
	\begin{align}
		\begin{aligned} \beta_{BH}^{-1} =  T & = \frac{1}{2 \pi  r_+} + \frac{r_+}{\pi  L^2} + \frac{\alpha '}{L^2} \left( \frac{1}{3 \pi r_+^3 L^2}\right) \Big( 4 \left(5 \beta _1-\beta _2+\beta _3\right) r_+^4 -6 \beta _2 L^2 \left(L^2+2 r_+^2\right) \Big) \\
			& +\left(\frac{\alpha '}{L^2} \right)^2 \left( \frac{1}{18 \pi  L^2 r_+^5} \right)\bigg[ 12 \alpha _1 \left(-21 L^6+18 L^4 r_+^2+99 L^2 r_+^4+62 r_+^6\right) \\
			& +9 \alpha _2 \left(L^6-6 L^4 r_+^2-15 L^2 r_+^4-6 r_+^6\right) +24 \alpha _3 \left(-3 L^6+6 L^4 r_+^2+21 L^2 r_+^4+14 r_+^6\right) \\
			& -24 \alpha _4 \left(9 L^6-30 L^4 r_+^2-87 L^2 r_+^4-58 r_+^6\right)  +32 r_+^6 \left(3 \alpha _5+3 \alpha _6+15 \alpha _8+75 \alpha _9\right)\\
			& +16 \beta _2^2 \left(36 L^6+36 L^4 r_+^2-39 L^2 r_+^4-38 r_+^6 \right) +64 r_+^6\left(5 \beta _1+\beta _3\right){}^2 +16 \beta _2 \left( 9 \beta _1 L^6 \right. \\ 
			& \left. +9 \beta _3 L^6 -18 \left(5 \beta _1+\beta _3\right) L^4 r_+^2 -3 \left(79 \beta _1+23 \beta _3\right) L^2 r_+^4 -2 \left(59 \beta _1+19 \beta _3\right) r_+^6\right)\bigg]. \end{aligned}
	\end{align}
	The temperature of the thermal adS can be fixed using equation \eqref{eq:betarel}. For 6-derivative it becomes
	\begin{equation}\label{eq:bads6der}
		\begin{aligned}
			\beta _{\text{AdS}} = & \beta _{\text{BH}} \bigg[ 1-\frac{L^2 r_0}{2 R_c^4} - \frac{\alpha^\prime}{L^2} \bigg( \frac{\left(10 \beta _1+\beta _2+2 \beta _3\right) L^2 r_0}{3 R_c^4} \bigg)\\
			& + \frac{\left(\alpha^\prime/L^2\right)^2}{18 R_c^4} \bigg( L^2 r_0 \Big(3 \left(4 \alpha _1+3 \alpha _2+8 \alpha _3+40 \alpha _4+16 \alpha _5+16 \alpha _6 \right.  \\
			& \qquad \qquad \left. +80 \alpha _8+400 \alpha _9\right)+4 \left(10 \beta _1+\beta _2+2 \beta _3\right){}^2\Big) \bigg)\bigg].
		\end{aligned}
	\end{equation}
	
	Following the background subtraction method the regularised on-shall Euclidean action is given by
	\begin{equation}
		\begin{aligned}
			{I_E}^{OS} = & -\tfrac{\beta_{BH} \Omega_3 }{16 \pi G} \bigg[r_+^2\left(1- \tfrac{r_+^2}{L^2}\right) + \tfrac{\alpha^\prime}{L^2} \bigg( -10 \beta _2 L^2-8 \left(5 \beta _1+3 \beta _2+\beta _3\right) r_+^2 \\
			& +\tfrac{20}{3 L^2} \left(5 \beta _1-\beta _2+\beta _3\right) r_+^4 \bigg) + \left( \tfrac{\alpha^\prime}{L^2} \right)^2  \bigg( \tfrac{2 L^4}{r_+^2} \left( 16 \alpha _1-7 \alpha _2 -8\alpha _3+24 \alpha _4)\right) \\
			&- \tfrac{32 L^4}{r_+^2} \beta _2 \left(\beta _1+\beta _2+\beta _3\right) -L^2 \left(8 \left(5 \alpha _3-25 \alpha _4-16 \beta _2^2\right)+12 \alpha _1+51 \alpha _2\right) \\
			& -\tfrac{r_+^4}{3 L^2}\left(268 \alpha _1+99 \alpha _2-56 \alpha _3+88 \alpha _4\right) + \tfrac{16 r_+^4}{9 L^2} (250 \beta _1^2+11 \beta _2 \beta _1+100 \beta _3 \beta _1 \\
			& +103 \beta _2^2+10 \beta _3^2+31 \beta _2 \beta _3) - r_+^2 \left(108 \alpha _1+51 \alpha _2+56 \alpha _3-376 \alpha _4\right)\\
			& -\tfrac{16 r_+^2}{3} \left(50 \beta _1^2+17 \beta _2 \beta _1+20 \beta _3 \beta _1-63 \beta _2^2+2 \beta _3^2-11 \beta _2 \beta _3\right) \bigg) \bigg]\,.
		\end{aligned}
	\end{equation}
	
	The ADM mass of a black hole can be computed from the free energy \eqref{freeE} using the definition \eqref{energy} 
	\begin{equation}\label{bhmass}
		\begin{aligned}
			M & = \frac{3 \Omega_3}{16 \pi G} \bigg[ r_+^2  + \frac{r_+^4}{L^2} + \frac{(\alpha '/L^2)}{3 L^2} \Big( 6 \beta _2 L^4-4 \left(5 \beta _1-\beta _2+\beta _3\right) r_+^2 \left(6 L^2+5 r_+^2\right) \Big) \\
			&  + \frac{(\alpha '/L^2)^2}{9 r_+^2 L^2} \bigg( 12 \alpha _1 \big(48 L^6+189 L^4 r_+^2+207 L^2 r_+^4+67 r_+^6\big)+9 \alpha _2 \big(2 L^6+15 L^4 r_+^2 \\
			& +45 L^2 r_+^4+33 r_+^6\big) +24 \alpha _3 \big(6 L^6+21 L^4 r_+^2+21 L^2 r_+^4+7 r_+^6\big)+24 \alpha _4 \big(18 L^6 \\
			& +57 L^4 r_+^2+45 L^2 r_+^4+11 r_+^6\big) +48 \alpha _5 r_+^4 \left(9 L^2+10 r_+^2\right)+48 \alpha _6 r_+^4 \left(9 L^2+10 r_+^2\right)\\
			& +240 \alpha _8 \left(9 L^2 r_+^4+10 r_+^6\right)+1200 \alpha _9 \left(9 L^2 r_+^4+10 r_+^6\right) -288 \beta _2\left(\beta _1+\beta _3\right) L^6 \\
			&  -32 \left(5 \beta _1+\beta _3\right){}^2 r_+^4 \left(3 L^2+5 r_+^2\right)-16 \beta _2^2 \left(54 L^6+216 L^4 r_+^2+267 L^2 r_+^4+103 r_+^6\right) \\
			& -16 \beta _2 \left(+72 \left(\beta _1+\beta _3\right) L^4 r_+^2+3 \left(25 \beta _1+29 \beta _3\right) L^2 r_+^4+\left(11 \beta _1+31 \beta _3\right) r_+^6\right) \bigg)\bigg]\,.
		\end{aligned}
	\end{equation}
	The entropy of the black hole can be computed using \eqref{entropy}
	\begin{equation}
		\begin{aligned}
			S = \frac{\Omega_3 r_+^3}{4 G} & \bigg[1 + \frac{(\alpha '/L^2)}{r_+^2} \left( 12 \beta _2 L^2 -8 \left(5 \beta _1-\beta _2+\beta _3\right) r_+^2 \right)\\
			& +\frac{(\alpha '/L^2)^2}{ 3 r_+^4} \Big( 27 \alpha _2 \left(L^2+2 r_+^2\right){}^2+\alpha _1 \left(3 L^2+2 r_+^2\right){}^2+72 \alpha _3 L^4 \\
			& \quad+72 \alpha _4 \left(3 L^4-4 L^2 r_+^2-2 r_+^4\right) +16 \left(9 \alpha _5+9 \alpha _6+45 \alpha _8+225 \alpha _9\right) r_+^4 \\
			& \quad-16 \beta _2^2 \left(27 L^4+54 L^2 r_+^2+26 r_+^4\right) -16 \beta _2 \left(9 \left(\beta _1+\beta _3\right) L^4 \right. \\
			& \quad \left. + 18 \left(\beta _1+\beta _3\right) L^2 r_+^2 + 4 \left(\beta _1+2 \beta _3\right) r_+^4\right)-32 \left(5 \beta _1+\beta _3\right){}^2 r_+^4 \Big)  \bigg]\,,
		\end{aligned}
	\end{equation}
	this expression is identical to what we computed from Walds's approach. In the main text we use these expressions by setting all the coefficients to zero except $\beta_2$, $\alpha_1$ and $\alpha_2$.

	\subsection{\texorpdfstring{$\mathcal{W}^4$}{}  gravity}
	Here we present the Euclidean computation for $AdS_5$ black hole in type IIB string theory with $\mathcal{W}^4$ term. The on-shall action can be depicted by
	\begin{equation}
		\begin{aligned}
			{I_E}^{\text{OS}} = \beta _{\text{BH}} \frac{\pi  r_+^2 \left(L^2-r_+^2\right)}{8 G L^2} \left( 1 + \gamma e^{-\frac{3}{2} \vphizero } \frac{5 \left(L^2-15 r_+^2\right) \left(L^2+r_+^2\right){}^3}{L^6 r_+^6 \left(L^2-r_+^2\right)} \right) \,,
		\end{aligned}
	\end{equation}
	Thus \eqref{freeE} implies the free energy of the black hole is
	\begin{equation}
		\begin{aligned}
			F = \frac{\pi  r_+^2 \left(L^2-r_+^2\right)}{8 G L^2} \left( 1 + \gamma  e^{-\frac{3}{2} \vphizero } \frac{5 \left(L^2-15 r_+^2\right) \left(L^2+r_+^2\right){}^3}{L^6 r_+^6 \left(L^2-r_+^2\right)} \right)
		\end{aligned}
	\end{equation}
	and the internal energy or ADM mass and the entropy of black hole is $ S = \frac{\partial F}{\partial T} $. The final expression is given by
	\begin{equation}
		\frac{\partial {I_E}^{\text{OS}}}{\partial \beta} \implies M = \frac{3 \pi  r_+^2 \left(L^2+r_+^2\right)}{8 G L^2} \left( 1 + \gamma e^{-\frac{3}{2} \vphizero } \frac{5 \left(L^2+r_+^2\right){}^2 \left(7 L^2+15 r_+^2\right)}{L^6 r_+^6} \right)\,,
	\end{equation}
	\begin{equation}
		\beta \frac{\partial I_E}{\partial \beta} - I_E \implies S = \frac{\pi^2 {r_+}^3}{2 G} \left( 1 + \gamma e^{-\frac{3}{2} \vphizero} \frac{60 \left(L^2+r_+^2\right){}^3}{L^6 r_+^6} \right)\,.
	\end{equation}

	\bibliographystyle{unsrt}
	\bibliography{ccref.bib}
	

\end{document}